\documentclass{icrc}

\usepackage{times}
\usepackage{graphicx} 

\begin{document}

\title{Observations of Mkn 421 using Pachmarhi Array of \v Cerenkov
Telescopes}
\author[1]{P. N. Bhat, B. S. Acharya, V. R. Chitnis, P. Majumdar, M. A. Rahman,
B. B. Singh and P. R. Vishwanath}
\affil[1]{Tata Institute of Fundamental Research, Colaba, Mumbai, 400005, India}

\correspondence{pnbhat@tifr.res.in}

\runninghead{P. N. Bhat {\it et al.}: PACT Observations of MKN 421 flares}
\firstpage{1}
\pubyear{2001}


\maketitle

\begin{abstract}
Pachmarhi Array of  \v Cerenkov Telescopes (PACT), based on
wavefront sampling technique, has been used for detecting TeV
gamma rays from galactic and extra-galactic $\gamma $-ray sources.
The Blazar, Mkn 421 was one such extra-galactic source observed during the
winter nights of
2000 and 2001.  We have carried out a preliminary analysis of the data taken
during the nights of January, 2000 and 2001. Results show a  significant gamma
ray signal from
this source during both these periods above a threshold energy of 900 GeV.
The source was contemporaneously observed by CAT imaging telescope during the
first episode of January 2000 while HEGRA CT1 was observing the source during
the second episode. Both these observations have detected variable $\gamma $-ray
emission this source and they reported that it was flaring during both these
periods.  The light curve in the TeV gamma ray range derived from the first PACT
observations during both these episodes is in agreement
with that reported by other experiments. The analysis procedure and the
preliminary results will be presented and discussed.
\end{abstract}

\section{Introduction}
Mkn 421 is an X-ray selected closest (z=0.03) BL Lacertae object exhibiting
extreme variability in VHE $\gamma -$ray emission. This is the first
extra-galactic object discovered at TeV energies [\cite{pu92}].
Variability time scale as low
as 15 minutes observed from this source [\cite{ga96}] implies a
compact emission region of dimension $R$ less than $10^{-4}$ pc which is only
an order of magnitude larger than the radius of the event horizon of a $10^8$
solar mass black hole. Correlations observed between X-rays and TeV
$\gamma -$rays [\cite{ma95,bu96,ca97}
are most easily explained as emissions at both
wavelengths produced by the same population of high energy electrons.
However models which produce $\gamma -$rays primarily through proton
interactions [\cite{ma93}] can also explain the observations so far. More
observations are still needed to understand the nature of the proginitors of
TeV $\gamma -$rays.

\section{Observations}

PACT is situated in the central Indian hill station Pachmarhi (longitude:
78$^{\circ}$ 26$^{\prime}$ E, latitude: 22$^{\circ}$ 28$^{\prime} N$ and
altitude: 1075 $m$). It consists of an array of \v Cerenkov detectors, each of
area 4.35 $m^2$,  deployed in the form of a rectangular array. The detectors in
the E-W direction have a separation of 25 $m$ and in the N-S direction
separation is 20 $m$. Each telescope consists of 7 parabolic reflectors mounted
para-axially on a single equatorial mount [\cite{bh98,bh01}].

The array has been divided into 4 sectors of six telescopes each and each sector
has its own data acquisition system which are networked with the master data
acquisition system. The analog signals from the seven individual mirrors of a
telescope are added in phase to generate a `royal sum'. The trigger is
generated when any 4 out of 6 royal sums are present in any sector. Following
an event trigger the TDC and the ADC informations from peripheral mirrors of
each telescopes, the UTC from a real time clock and the latch information are
recorded. The master recording system at the main control room records the TDC
information from all royal sums [\cite{up01}].

PACT has become completely operational since December 2000. Two of the four
sectors (consisting of a total of 12 telescopes) were ready about an year ago
when the
science observations were started. Various celestial point sources of
$TeV~\gamma $-rays have been observed. Among these are the galactic sources
like the Crab Nebula, Geminga and extra-galactic sources like Mkn 421 and
Mkn 501. We have detected positive signal from the Crab Nebula [\cite{bh01}]
and the detected flux is consistent with the expected sensitivity of PACT.

The flaring Blazar Mkn 421 has been observed by PACT (using sectors 3 \& 4 only
during
January 2000 flare and all the 4 sectors during the January 2001 flare). We
have a total of 73.3 hrs of ON-source date and 50.6 hrs of OFF-source data
during the 2000 observing season while we have logged 45.6 hrs of ON-source
data and 39.7 hrs of OFF-source data until the end of February during 2001
observing season.

\section{Analysis}

From the vast amount of data we have carried out only a
preliminary analysis of the January 2000 and 2001 data because of the flaring
activity
reported by other groups [\cite{gd00,br01}]

The arrival direction of each event with more than 8 telescopes
participating, has been estimated using a plane front approximation. In this
preliminary analysis only the telescope TDC values are used. The TDC information
from individual mirrors in a telescope are not used. The mean value of the
$\chi ^2$ values of the plane front fit
to the TDC data is around 1. We rejected events with $\chi ^2$ values more
than 1 standard deviation above the mean. The off-source data covering the
same hour angle range as the corresponding on-source run  is
taken on the same night  as the source run  or    the following night. On \& off
-source data segments are  of same duration. The space angle between the
direction of the primary estimated as mentioned above and the source or the
off-source direction as the case may be is estimated for each accepted event.
The space angle distributions are generated both for on \& off source data.
The two distributions are normalized
from the number of counts in the space angle range $3^\circ-5^\circ$ and the
off-source distribution is subtracted from the on-source distribution to
estimate the $\gamma $-ray signal from the source. Different space angular
ranges were used for normalization and the RMS value of the estimated signal is
used as a measure of systematic error on the signal. It has been found that
the signal strength is not a sensitive function of space angle range used for
normalization. In all cases the space angle distributions of the signal are
similar,
showing excess close to zero and reaching zero counts with increasing space
 
angle while the FWHM of this distribution is consistent with the angular
resolution [\cite{ma01}].

\section{Results}

\begin{figure}[t]
\vspace*{2.0mm} 
\includegraphics[width=8.3cm]{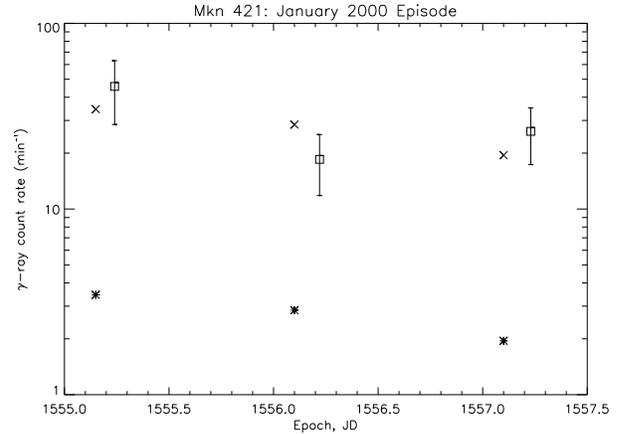} 
\caption{The $\gamma $-ray count rate from Mkn 421 as a function of epoch (MJD)
during January 2000. The count rate observed by CAT during January 2000 are
shown as asterisks. These are re-plotted after multiplying by 10 
(shown as crosses) for comparison of the variability in the two independent
observations.
}
\end{figure}

\begin{figure}[t]
\vspace*{2.0mm} 
\includegraphics[width=8.3cm]{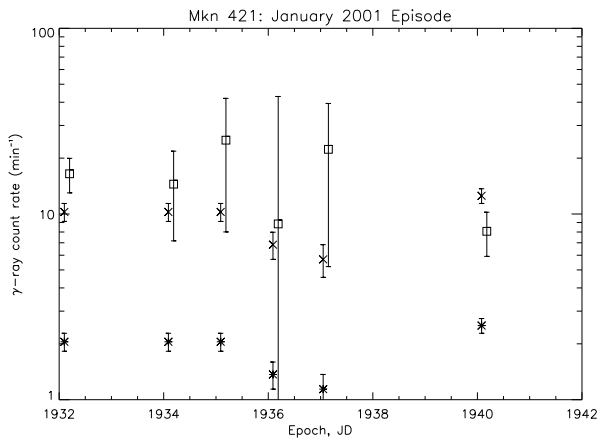} 
\caption{The $\gamma $-ray count rate from Mkn 421 as a function of epoch (MJD)
January 2001. The count rate observed by observed HEGRA CT1 during January 2001
are shown as asterisks. These are re-plotted after multiplying by 5 (shown as
crosses) for comparison of the variability in the two independent observations.
}
\end{figure}

Figure 1  shows a plot of the $\gamma $-ray count rate from Mkn 421 during each
night of observation on January 11, 12 \& 13, 2000 as a function of epoch. These
are shown as open squares and the error bars include systematic errors as
mentioned above. Also
shown in the same figure are the counting rates above 250 GeV from the CAT
imaging telescope (shown as asterisks, \cite{gd00}). These
rates are multiplied by a factor
of 10 and re-plotted (shown as diamonds) for better comparison of the
variability of the two independent results. It can be seen that the two light
curves, one with primary $\gamma $-ray $>$ 250 GeV
[\cite{pi99}] and the other with primary $\gamma $-ray $>$ 900 GeV
are very similar.

Similarly, Figure 2 shows the nightly rates of TeV $\gamma $-rays from Mkn 421
above 900 GeV from PACT observations during January 2001. Also shown in
the same plot are the $\gamma $-ray ($>$ 500 GeV) count rate contemporaneously
detected by HEGRA CT1 [\cite{br01}]. Figure 1 also shows a
re-plot of  the HEGRA CT1 rates multiplied by 5 as diamonds to
enable one to compare the counting rate variability in the two independent
observations. During both the episodes there is a good agreement between the
two light curves despite different energy thresholds.

\section{Discussions}

It may be noted that this is only a preliminary result and hence we have
not estimated the TeV $\gamma $-ray flux from this source during the flaring
activity. The enhanced counting rate from PACT as compared that from imaging
telescopes, is attributed to its increased collection area.

We have carried several checks to improve the confidence level of the signal
seen by PACT.
A positive signal could result, for example, if the space angle distribution of
background region has some systematic problem. Hence we compared the
background data taken during two different nights and after subtracting one
from the
other no positive excess was detected in any space angle range. In addition,
we compared an ON-source data with several nights' background data separately
and estimated the signal in each case. Table 1 shows the systematic variation in
the signal strength resulting from the systematic variation in the
background data taken during different nights. In addition, each on/off source
data set has been subdivided
as per the number of telescopes that have participated during the event. Each
subset was independently analyzed and the signals estimated. The variation of
signal strengths from segment to segment were consistent within the expected
variation in each run.
\vskip 2cm
\begin{table}
\caption{A table showing the consistency of signal strength with different
backgrounds}
\vskip 0.3cm
\begin{tabular}{lllllll}
\hline
\hline
\# of ON & \# of OFF   & Excess & Sigma &Durn & $\gamma $-ray \\
Source & Source & (S-B) & ${(S-B)}\over{\sqrt{S+2B}}$ & ($mins.$)& rate\\
events & events &      &                             &          & ($min^{-1}$)\\
\hline
14543 & 11508 & 3035 & 15.7 & 136.6 & 22.2$\pm$4.7\\
\hline
 3311 &  2104 & 1207 & 16.4 &  68.9 & 17.5$\pm$4.2\\
\hline
 7180 &  5771 & 1409 & 12.4 &  68.4 & 20.6$\pm$4.5\\
\hline
 6604 &  4654 & 1950 & 15.5 &  62.2 & 31.4$\pm$5.6\\
\hline
\end{tabular}
\end{table}
\begin{figure}[t]
\vspace*{2.0mm} 
\includegraphics[width=8.3cm]{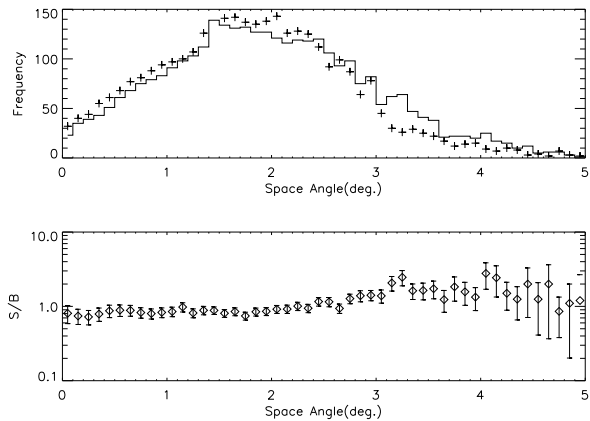} 
\caption{The space angle distribution of two sets of background data taken on  
two different nights. The two data sets span the same hour angle range 
(upper panel). The lower panel shows the ratio of the two distributions 
as a function of space angle showing that when two background data-sets are 
compared no signal is seen adding credibility to the signal from the source
reported here.
}
\end{figure}

Figure 3 shows the the space angle distribution of events when two sets of
background data are compared. The space angle distribution of the ratio's show
no evidence of any signal showing that the systematic variations, if any,  in
the background data cannot masquerade as a $\gamma $-ray signal.
 
\section{Conclusions}

A variable TeV $\gamma $-ray signal above 900 GeV has been detected by PACT for
the first time during
both the episodes in January 200 and January 2001. The variability measurements
are compatible with those independently seen by other atmospheric
\v Cerenkov telescopes.
The fourth run taken in January 2000 did not show any signal once again
confirming that PACT analysis is working. However there are more data from
this source to be analyzed.


\begin{thebibliography}{99}

\bibitem[Bhat (1998)]{bh98}
Bhat, P. N., 1998, {\it ``High Energy Astronomy \& Astrophysics''}
, Proc. of the Int. Colloquium to commemorate the Golden Jubilee year of
 Tata Institute of Fundamental Research, Ed: P. C. Agrawal and P. R.
Vishwanath, University Press, 370

\bibitem[Bhat (2001)]{bh01}
 Bhat, P. N., 2001, {\it Wavefront Sampling Technique: VHE $\gamma -$ray
 Experiments in India}, {\it ibid.}

\bibitem[Boerst \& Remillard (2001)]{br01}
 Boerst, H. G. and Remillard, R. (for HEGRA Collaboration), 2001, IAU Circular
No. 7568 and http://www-hegra.desy.de/mrk-421/

\bibitem[Buckley {\it et al.} (1996)]{bu96}
 Buckley, J. H., {\it et al.}, 1996, {\it Astrophys. J}, 472, L319.

\bibitem[Catanese {\it et al.} (1997)]{ca97}
 Catanese, M., 1997, {\it et al.}, 1997, {\it ApJ}, 487, L143.

\bibitem[Gaidos {\it et al.} (1996)]{ga96}
 Gaidos, J. A. {\it et al.}, 1996, {\it Nature}, 383, 319.

\bibitem[Gouiffes \& Degrange (2000)]{gd00}
 Gouiffes, C. and Degrange, B. (for CAT Collaboration), 2000, IAU Circular No.
 7345.

\bibitem[Macomb {\it et al.} (1995)]{ma95}
 Macomb, D. J., {\it et al.}, 1995, {\it Astrophys. J}, 449, L99.

\bibitem[Mannheim (1993)]{ma93}
 Mannheim, K., 1993, {it Astron. \& Astrophys.}, 269, 67

\bibitem[Majumdar {\it et al.} (2001)]{ma01}
 Majumdar, P. {\it et al.}, 2001, {\it Angular Resolution of the Pachmarhi
Array of \v Cerenkov Telescopes},{\it Wavefront Sampling Technique: VHE $\gamma -$ray
 Experiments in India}, {\it ibid.}

\bibitem[Piron {\it et al.} (1999)]{pi99}
 Piron, F. for the CAT Collaboration, 1999, {\it 26th International Cosmic Ray
Conference, Utah}, OG 2.1.09; Astro-ph/9906102.

\bibitem[Punch {\it et al.} (1992)]{pu92}
 Punch, M.  {\it et al.}, 1992, {\it Nature}, 358, 477.

\bibitem[Upadhya {\it et al.} (2001)]{up01}
 Upadhya, S. S. {\it et al.}, 2001, {\it  Distributed Data Acquisition System
for Pachmarhi Array of \v Cerenkov Telescopes}, {\it Wavefront Sampling Technique: VHE $\gamma -$ray
 Experiments in India}, Proc. of {\it Int. Symp. on Gamma-ray Astrophysics
through Multiwavelength Expts}: GAME-2001, March 8-10, 2001, Mt. Abu, India
(Ed. R. K. Kaul and C. L. Kaul)


\end{thebibliography}
\end{document}